\documentclass[twocolumn,nofootinbib]{revtex4-1}
\usepackage[utf8]{inputenc}
\usepackage{amsmath}
\usepackage{amssymb}
\usepackage{amsfonts}
\usepackage{graphicx}
\usepackage{color}
\usepackage{amsbsy}
\usepackage{url}
\usepackage{enumerate}
\usepackage[vcentermath,enableskew]{youngtab}
\usepackage{verbatim}
\usepackage[normalem]{ulem}
\usepackage{cancel}
\usepackage{bm,bbm}
\usepackage{slashed}

\newcommand{\refer}[1]{(\ref{#1})}

\newcommand{\comm}[2]{\bigl[ #1, #2 \bigr]}

\newcommand{\Tr}{\mathrm{Tr}}

\newcommand{\sech}{\mbox{sech}}

\def\be{\begin{eqnarray}}
\def\ee{\end{eqnarray}}
\def\p{\partial}

\begin{document}


\title{
Grand Unified  Brane World Scenario
}

\author{Masato Arai}
\email{arai(at)sci.kj.yamagata-u.ac.jp}
\affiliation{Faculty of Science, Yamagata University, 
Kojirakawa-machi 1-4-12, Yamagata,
Yamagata 990-8560, Japan 
}

\author{Filip Blaschke}
\email{filip.blaschke(at)fpf.slu.cz}
\affiliation{
Faculty of Philosophy and Science, Silesian University in Opava, Bezru\v{c}ovo n\'am. 1150/13, 746~01 Opava, Czech Republic, \\
Institute of Experimental and Applied Physics, Czech Technical University in Prague, Horsk\'a 3a/22, 128 00 Praha 2, Czech Republic
}

\author{Minoru Eto}
\email{meto(at)sci.kj.yamagata-u.ac.jp}
\affiliation{Department of Physics, Yamagata University, 
Kojirakawa-machi 1-4-12, Yamagata,
Yamagata 990-8560, Japan
}

\author{Norisuke Sakai}
\email{norisuke.sakai(at)gmail.com}
\affiliation{Department of Physics, and Research and 
Education Center for Natural Sciences, 
Keio University, 4-1-1 Hiyoshi, Yokohama, Kanagawa 223-8521, Japan, 
and iTHEMS, RIKEN,
2-1 Hirasawa, Wako, Saitama 351-0198, Japan
}

\begin{abstract}

We present a field theoretical model unifying  grand unified 
theory (GUT) and brane world scenario.
As a concrete example, we consider $SU(5)$ GUT in 4+1 dimensions 
where our $3+1$ dimensional spacetime spontaneously 
arises on five domain walls. 
A field-dependent gauge kinetic term is used to localize 
massless non-Abelian gauge fields on the domain walls and to assure 
the charge universality of matter fields.
We find the domain walls with the symmetry 
breaking $SU(5)\to SU(3)\times SU(2)\times U(1)$ as a global 
minimum and
all the undesirable moduli are stabilized with the mass scale of 
$M_{\rm GUT}$. 
Profiles of massless Standard Model particles  are determined as a 
consequence of wall dynamics.
The proton decay can be exponentially suppressed.

\end{abstract}

\maketitle

\newpage

%


\section{Introduction}
Grand unified theories (GUT)  \cite{Pati:1973uk, 
Georgi:1974sy, Georgi:1974yf,Witten:1981nf,Dimopoulos:1981zb,
Sakai:1981gr,Dimopoulos:1981yj} have various interesting features 
such as prediction of gauge coupling unification and explanation 
of the charge assignments of matters.
Theories with extra-dimensions give a solution of the gauge 
hierarchy problem in the framework such as the brane world 
scenario \cite{ArkaniHamed:1998rs,Antoniadis:1998ig, 
RS,RS2,Akama}. 
Combining these two theories gives rise to new possibilities 
for phenomenological model building.

An interesting example
is $SU(5)$ GUT in five dimensional 
space-time where the fifth dimension is {\it compactified} on an orbifold \cite{Kawamura:1999nj}. 
In this kind of model, several assumptions are made:
i) The fifth dimension is compactifed. 
ii) Two branes exist. iii) Matter fields are localized on one of two branes called an Infra-Red brane 
while gauge fields and Higgs fields propagate in the bulk. iv) Nontrivial $\mathbf{Z}_2$ parity
assignment for the fields is required.
This setup leads to breaking 
of $SU(5)$ gauge group to the Standard Model (SM) gauge group
and realizes chiral fermions on the Infra-Red brane.
Further studies along this direction have been done to reconsider 
traditional problems in GUT such as doublet-triplet Higgs 
splitting and proton decay which has not been observed so far
\cite{Kawamura:2000ev, Kawamura:2000ir, Altarelli:2001qj,
Hall:2001pg, Hebecker:2001wq}. 
A key feature in these models is gauge symmetry breaking via  
orbifold compactification. 
However, the origins of the nontrivial space-time geometry (only the fifth direction is compact) 
and the $\mathbf{Z}_2$ parity assignment has not been explained.
Moreover, both the presence of infinitely thin branes and the localization of matters on an Infra-Red brane are assumed
as an initial setup.  

These points, however, can be addressed
in {\it non-compact} five dimensional space-time.
A minimal assumption is a presence of discrete  vacua which
are exchanged by a discrete symmetry.
Spontaneous symmetry breaking of the discrete symmetry dynamically 
yields stable domain walls, which are solutions of equations of motion.
Thus, our four-dimensional world is dynamically realized as the domain walls with finite width in flat 5 dimensional spacetime.
Furthermore, they
automatically lead to localization of zero modes of matter fields in the bulk such as fermions 
and scalars \cite{Rubakov}.
It provides a dynamical realization of four-dimensional space-time, namely,
the brane world scenario.

An $SU(5)$ GUT model in the five-dimensional space-time with a 
domain wall background has been proposed in Ref. \cite{Volkas}, 
and has been studied extensively in 
\cite{Davidson:2007cf, Volkas2, Callen:2010mx, Callen:2012kd}. 
The model in \cite{Volkas} 
introduces three scalar fields: a singlet field $\eta$ ($\mathbf{1}$), an adjoint field $\chi$  ($\mathbf{24}$) and the Higgs field in the anti-fundamental representation $\Phi$ ($\mathbf{5}^*$). 
The scalar $\eta$ forms a domain wall on which a low-energy effective $3+1$-dimensional 
space-time is realized.
The other scalar field $\chi$ yields a background solution which breaks the $SU(5)$ gauge 
symmetry group down to the SM gauge group, while $\Phi$ is localized on the domain wall and gives the origin 
of the electroweak symmetry breaking.
Thus, each scalar field has an independent role. 
Their model successfully combines
 GUT and the brane world scenario through the domain wall and demonstrates the feasibility of this approach for an interesting model building.

The localization of gauge fields on the domain walls has been 
a long-standing problem in model building of the brane world 
scenario by topological solitons \cite{Dvali:2000rx, Kehagias:2000au, 
Dubovsky:2001pe, Ghoroku:2001zu,Akhmedov:2001ny, Kogan:2001wp, 
Abe:2002rj, Laine:2002rh, Maru:2003mx, Batell:2006dp, Guerrero:2009ac, 
Cruz:2010zz, Chumbes:2011zt, Germani:2011cv, Delsate:2011aa, 
Cruz:2012kd, Herrera-Aguilar:2014oua, Zhao:2014gka, Vaquera-Araujo:2014tia}. 
A popular resolution is the so called Dvali-Shifman mechanism \cite{Dvali:1996xe}.
Indeed, the SM gauge fields  in \cite{Volkas} are assumed to be localized 
due to the Dvali-Shifman mechanism.
For this mechanism to work, it needs the confinement in five-dimensional 
space-time, whose validity is far from being clear. 

It has been noted that the localization of gauge fields requires 
the confining phase rather than the Higgs phase in the bulk 
outside the domain wall \cite{Dvali:1996xe,ArkaniHamed:1998rs}. 
A classical realization of the confinement can be obtained by 
the position-dependent gauge coupling \cite{Kogut:1974sn,Fukuda:1977wj}, 
which is achieved by domain wall through the 
field-dependent gauge coupling function. This semi-classical 
mechanism was successfully applied to localize gauge fields 
on domain walls \cite{Ohta,Us1,Us2,1stpaper}.

In this work, we propose an alternative way to unify GUT 
and extra-dimensions, where the (effective) compactification, 
gauge symmetry breaking and especially localization of gauge fields are all tied to domain walls.
We also show that charge universality of matter fields holds 
and the proton decay is suppressed. 
Our model is an $SU(5)$ GUT in non-compact five-dimensional space-time. 
Our $3+1$ dimensional world emerges dynamically on domain walls. 
Having multiple domain walls has mainly three roles.
First is dynamical compactification of the fifth direction. 
The second role is that the SM chiral fermions are localized through 
the mechanism in \cite{Rubakov}. 
These are the conventional roles played by domain walls in previous studies. In addition,  
non-coincident positions of domain walls in the extra-dimension break $SU(5)$ gauge symmetry. 
Hence, the presence of domain walls in our model is essential not only for 
dynamical realization of the brane world but also for GUT scenario.
Rephrasing at more concrete level, we unify the roles of $\eta\, (\mathbf{1})$ and $\chi\, (\mathbf{24})$ of Ref. \cite{Volkas} into a single entity, which we call $T\, (\mathbf{1}+\mathbf{24})$. 
Peculiar point as a result of this unification is that the number of domain walls is equal to the rank of $SU(5)$. 
We emphasize that the GUT symmetry breaking and its stabilization are results of  
dynamics of these domain walls.
 
In the absence of the moduli-stabilizing potential, our model 
possesses five domain walls, whose positions are moduli. 
Gauge symmetry is fully preserved when all the five walls are 
coincident at the same position. 
When a group of three coincident walls and the other group of 
two coincident walls are located at two different points, the 
gauge symmetry is broken down to that of the SM.  
With a simple moduli-stabilizing potential, we show that our 
model has such 3-2 splitting pattern  
as the global minimum.

Our model also overcomes the problem of localizing 
gauge fields on the domain walls by adopting field-dependent 
gauge kinetic term \cite{Ohta, Us1,Us2}. 
As an advantage of using this mechanism, we can explicitly 
determine mode functions of massless gauge bosons for the SM 
gauge groups, as well as mode functions of the gauge fields 
corresponding to broken generators.
We show that these gauge bosons for broken generators become 
massive by the geometric Higgs mechanism \cite{1stpaper}. 
These massive gauge bosons are responsible for the proton decay.
Their mode functions and those of the matter fields allow us to 
show explicitly that the proton decay can be highly suppressed. 
This suppression arises due to the small overlap between the 
mode functions of the matter fields and the massive gauge fields. 
On the other hand, our localization mechanism assures charge 
universality of matter fields for unbroken gauge generators 
by preserving the $3+1$-dimensional gauge invariance. 

The organization of this paper is as follows. In Section \ref{sec:wall}, 
we present our model, domain walls and fluctuations without 
moduli-stabilizating potential. In Section \ref{sec:moduli_stabilization}, 
we introduce the moduli-stabilizating potential 
and find the global minimum using gradient flow. 
In section \ref{sec:quark_lepton}, 
one generation of chiral quarks and leptons are obtained. 
In section \ref{sec:baryon_number_viol}, baryon number violating 
processes are discussed. 
Section \ref{sec:conclusion} is devoted to Conclusion.

\section{Domain walls and gauge fields}
\label{sec:wall}
We extend the minimal $SU(5)$ GUT in $3+1$ dimensions to $4+1$ 
dimensions. 
In addition to fermionic matters $\Psi_{\bar 5}$ in an anti-fundamental 
representation and $\Psi_{10}$ in an anti-symmetric representation, 
we have two scalar fields represented as 5 by 5 matrices
\begin{equation}
T \equiv \hat T +\mathbf{1}_5\frac{T^0}{5}, \quad 
S \equiv \hat S +\mathbf{1}_5\frac{S^0}{5},
\end{equation} 
where $\hat T$ and $\hat S$ are in the adjoint representation 
while $T_0$ and $S_0$ are singlets.
The minimal GUT in $3+1$ dimensions needs only $\hat T$ to break 
$SU(5)$ gauge symmetry.
In our model, $\hat T$ plays the same role, but together with 
$T_0$, it provides domain walls and traps the chiral fermions 
on them. 
On the other hand, the role of $S$ is to localize the 
massless gauge fields.
Our  Lagrangian consists of three parts 
\begin{equation}
{\mathcal L} = 
{\mathcal L}_{\rm B}+{\mathcal L}_{\rm OS}+{\mathcal L}_{\rm F}. 
\end{equation}
The first term ${\mathcal L}_{\rm B}$ gives the bosonic 
Lagrangian to form domain walls 
\begin{equation}
{\mathcal L}_{\rm B} = \Tr\bigl[D_M T D^M T
+D_M S D^M S  -V_0 - V_1\bigr]
\end{equation}
 with the potential 
\begin{align}
 \label{eq:volpot}
V_0 &= \lambda^2\bigl(v^2 -T^2
-S^2\bigr)^2+\Omega^2S^2 - \xi \comm{T}{S}^2\,,\\
V_1 
&= -\mu^2 \hat T^2 + \alpha \hat T^4\,,
\label{eq:V1}
\end{align}
where metric is $\eta_{MN} = {\rm diag}(+,-,-,-,-)$, and 
$\Omega$, $\lambda$ and $\xi$ are mass and coupling constants, 
while the covariant derivatives are defined by 
\begin{equation}
D_M T = \partial_M T + i\comm{A_M}{T},  
\end{equation}
and similarly for $S$. 
We choose the  potential \refer{eq:volpot}  
to be simple to ensure analytic solutions if $V_1$ is 
absent \cite{1stpaper}. 
The role of $V_1$ in (\ref{eq:V1}) with $\alpha,\mu^2 > 0$ is 
to stabilize undesirable moduli of the domain wall solutions. 
The field-dependent gauge kinetic function 
\cite{Ohta,Us1,Us2} is given in ${\mathcal L}_{\rm OS}$ as
\begin{equation}\label{eq:ohta}
{\mathcal L}_{\rm OS}  = -  \Tr\bigl[a S^2 G_{MN}G^{MN}\bigr]\,,
\end{equation} 
with 
\begin{equation}
G_{MN} = \partial_MA_N-\partial_N A_M +i \comm{A_M}{A_N}.
\end{equation}
The fermionic part ${\mathcal L}_{\rm F}$ is given by 
\be
{\mathcal L}_{\rm F} &=& i \bar{\Psi}_{\bar 5}\Gamma^M D_M\Psi_{\bar 5}
+ i \Tr\bigl[\bar{\Psi}_{10}\Gamma^MD_M \Psi_{10}\bigr] \nonumber\\
&&+\,
h\bar{\Psi}_{\bar 5}T^t\Psi_{\bar 5}+ \tilde h \Tr
\bigl[\bar{\Psi}_{10}T\Psi_{10}\bigr]\,, 
\label{eq:L_F}
\ee
where $h$ and $\tilde h$ are Yukawa couplings, while the covariant 
derivatives are 
\begin{equation}
D_M \Psi_{\bar 5} = \partial_M \Psi_{\bar 5}-i A_M^* \Psi_{\bar 5}, 
\end{equation}
\begin{equation}
 D_M \Psi_{10} = \partial_M \Psi_{10} + i A_M \Psi_{10}
+i \Psi_{10}A_M^{t}. 
\end{equation}

Let us first consider the case without the moduli-stabilizing 
potential : $V_1 = 0$. 
The potential $V_0$ gives the multiple degenerate vacua 
\begin{equation}
T = \pm v \Lambda, \quad S = \mathbf{0}_5, 
\quad \Lambda^2 = \mathbf{1}_5. 
\end{equation} 
Up to symmetry transformations, we can label the vacua by 
the number of $\pm1$ eigenvalues of $\Lambda$. 
For example, we refer to the vacuum with $\Lambda = {\rm diag}(1,1,1,-1,-1)$ 
as $\left<3,2\right>$. 
Clearly, $\Lambda$ determines the breaking pattern of $SU(5)$. 
In the vacua, the components in $T$ except for those eaten by 
the gauge fields have masses 
$m_T = \sqrt{2}\, \lambda v$, and all components in $S$ have 
$m_S = \Omega$. 
The fermion masses are $m_{\bar 5}= hv$ and $m_{10}= \tilde h v$. 
We assume all these masses to be the same order or larger than 
 $M_{\rm GUT}$.

Since all these vacua are degenerate and discrete, static and 
stable domain walls exist. 
We choose the $SU(5)$ unbroken vacua as the boundary condition 
\begin{equation}
T \to \pm v {\bf 1}_5, \quad 
S \to 0, 
\label{eq:boundary_cond}
\end{equation}
at $x^4=y \to \pm \infty$, and assume 
\begin{equation}
\bar v^2 \equiv v^2 -\frac{\Omega^2}{\lambda^2} > 0. 
\end{equation}
Then we find exact domain walls solutions
\begin{equation}\label{eq:gensol3}
T  = v\tanh\Omega\bigl( y \mathbf{1}_5-Y\bigr)\,, \ 
S  = \bar v\,\sech\, \Omega \bigl( y\mathbf{1}_5-Y\bigr)\,,
\end{equation}
with a $5\times 5$ Hermitian matrix $Y$ containing all 
the parameters i.e. moduli of the solution. 
Without loss of generality, we can diagonalize $Y$.
Depending on the number of coincident walls and the ordering 
of their positions on $y$, we find ten qualitatively different patterns. 
Phenomenologically most interesting one is the 3-2 splitting 
configuration with 
\begin{equation}
Y = {\rm diag}({\cal Y}_3,{\cal Y}_3,{\cal Y}_3,{\cal Y}_2,{\cal Y}_2), 
\end{equation}
where three walls are located at $y = {\cal Y}_3$ and the 
remaining two are at $y ={\cal Y}_2$, interpolating 
the left-most vacuum $\left<0,5\right>$, the middle vacuum 
$\left<3,2\right>$, and the right-most vacuum $\left<5,0\right>$. 
The $SU(5)$ is broken to $ G_{\rm SM} = SU(3)\times SU(2) \times U(1)_Y$ 
in the middle $\left<3,2\right>$ vacuum. 

To identify the four dimensional effective fields and their 
spectra around this configuration, let us consider small 
fluctuations \cite{1stpaper}
\be
T &=& 
T_0(y)
+ \left(\begin{
matrix}
t_3(x^\mu,y) & \tilde t(x^\mu,y)\\
\tilde t(x^\mu,y)^\dagger & t_2(x^\mu,y)
\end{
matrix}
\right),
\label{eq:fluctuation}
\ee
where $T_0$ stands for the background solution, $t_3$ ($t_2$) 
is $3\times3$ ($2\times 2$) Hermitian matrix, and $\tilde t$ 
is 3 by 2 complex matrix.
We consider fluctuations for $S$ similarly 
\be
S &=& 
S_0(y)
+ \left(\begin{
matrix}
s_3(x^\mu,y) & \tilde s(x^\mu,y)\\
\tilde s(x^\mu,y)^\dagger & s_2(x^\mu,y)
\end{
matrix}
\right).
\label{eq:fluctuation_S}
\ee
The fluctuations of gauge fields are given as
\begin{equation}
A_\mu = \left(\begin{
matrix}
a_{3\mu} & b_\mu \\
b_\mu^{\dagger} & a_{2\mu}
\end{
matrix}\right)
+ a_{1\mu} \sqrt{3/5}\left(
\begin{smallmatrix}
\frac{1}{3}\mathbf{1}_3 & 0 \\
0 & - \frac{1}{2}\mathbf{1}_2
\end{smallmatrix}
\right)\,,
\label{eq:vectorFluctuation}
\end{equation}
where $a_{2}^{\mu} (a_3^\mu, a_1^\mu)$ corresponds to 
$SU(2), (SU(3), U(1))$ gauge field, and $b_\mu$ is the $3\times2$ 
complex matrix. 
It was found 
\cite{1stpaper} that the lightest modes in $t_{3} (t_{2})$ 
are massless, and those with the index 3 (2) are localized around 
the 3 (2) coincident walls at $y={\cal Y}_3$ ($y={\cal Y}_2$). 
Localization of the gauge fields is achieved by the field 
dependent gauge kinetic term (\ref{eq:ohta}).
Let us define 
\begin{equation}
\sigma_{2,3} = \bar v \sinh \Omega (y-{\cal Y}_{2,3}), 
\quad 
\sigma_1 = \sqrt{\frac{2\sigma_3^2+3\sigma_2^2}{5}}.
\end{equation} 
Then the fluctuations $a_{\alpha,\mu}$ ($\alpha = 1,2,3$) in 
the axial gauge $(a_{\alpha,y}=0)$ is expanded 
\be
a_{\alpha,\nu} = \sum_n \omega_{\alpha,\nu}^{(n)}(x) 
\frac{v_\alpha^{(n)}(y)}{\sigma_\alpha(y)},
\label{eq:wf_gauge}
\ee
with the four-dimensional effective vector fields $\omega_{\alpha,\nu}^{(n)}$. 
Here we are interested in divergence-free part 
$\p^\nu  a_{\alpha,\nu}
 = 0$.
The lowest mode was found \cite{1stpaper} to be massless and its 
mode function is flat 
\begin{equation}
\frac{v_\alpha^{(0)}(y)}{\sigma_\alpha(y)} = 1.
\end{equation}
Nevertheless, the gauge field zero modes are 
localized on domain walls thanks to the field dependent gauge kinetic 
function $aS^2$ in (\ref{eq:ohta}) which
provides an effective mode function 
\begin{equation}
\sqrt{a}\,\sigma_\alpha(y) \times 
\frac{v_\alpha^{(0)}(y)}{\sigma_\alpha(y)}
= \sqrt{a}\,\sigma_\alpha(y)\,.
\label{eq:flat_profile}
\end{equation}
Hence, the massless $SU(3)$ ($SU(2)$) gauge fields are localized 
around $y={\cal Y}_3$ ($y={\cal Y}_2$), and those of $U(1)_Y$ are 
around both $y={\cal Y}_2$ and ${\cal Y}_3$.
The dimensionless effective gauge couplings for  
$G_{\rm SM}$ gauge group are 
\begin{equation}
\frac{1}{g_3^2}  =  \frac{1}{g_2^2} = \frac{1}{g_1^2} = \frac{4a\bar v^2}{\Omega} 
\equiv \frac{1}{g^2}.
\end{equation}

The zero modes $\tilde t^{(0)}$ contained in $\tilde t$ 
(the off-diagonal component of $T$) in Eq.(\ref{eq:fluctuation}) and 
$\tilde s^{(0)}$ in $\tilde s$ in Eq.(\ref{eq:fluctuation_S})
are the would-be Nambu-Goldstone (NG) modes for the 
spontaneous breaking of $SU(5) \to G_{\rm SM}$, and are extended between 
the 3 coincident walls and the 2 coincident walls  
\be
\left(
\begin{
matrix}
\tilde t^{(0)}\\
\tilde s^{(0)}
\end{
matrix}
\right)
=  \left(
\begin{
matrix}
\tau_3 - \tau_2\\
\sigma_3 - \sigma_2
\end{
matrix}
\right)\tilde \eta^{(0)}(x)
\equiv \tilde{\mathbbm{u}}^{(0)}\tilde \eta^{(0)}(x)\,,
\label{eq:zero_t}
\ee
where $\tilde \eta^{(0)}(x)$ is a 3 by 2 complex matrix and 
\begin{equation}
\tau_{3,2} = v \tanh \Omega(y-{\cal Y}_{3,2}). 
\label{eq:tau_function}
\end{equation}
They are absorbed by 
the off-diagonal gauge fields $b_\mu$ in 
Eq.(\ref{eq:vectorFluctuation}) due to the geometric 
Higgs mechanism \cite{1stpaper}. 
Let us expand the divergence-free part of $b_\mu$ ($\partial^\mu b_\mu=0$) as 
\be
b_\mu = \sum_n \beta^{(n)}_\mu(x) {\gamma^{(n)}(y) \over \sigma_+(y)}\,.
\ee
with 
\begin{equation}
\sigma_+ = \sqrt{\sigma_2^2+\sigma_3^2}.
\end{equation}
Although it is difficult to derive the spectrum analytically we can obtain 
qualitative features \cite{1stpaper}.
When $L = |{\cal Y}_3 - {\cal Y}_2|$ vanishes, gauge field has 
zero mode (no symmetry breaking). 
Hence the mass of the lightest mode starts linearly for small 
$L$ as 
\begin{equation}
\tilde \mu_0 = \Omega L \sqrt{\frac{2v^2 + \bar v^2}{6a\bar v^2}} 
 \sim  \Omega L M_{\rm GUT} \ll M_{\rm GUT}
 \end{equation} 
 for 
$\Omega L \ll 1$. 
On the other hand, when $\Omega L \gtrsim 1$, $b_\mu$ is 
localized around the well-separated walls with the width 
$\Omega^{-1}$. 
Therefore, the mass becomes independent of $L$ as 
$\tilde \mu_0 = \sqrt{\Omega^2 + 1/a} \sim M_{\rm GUT}$. 
Apart from the moduli fields which will be discussed in the 
subsequent section, we find that all the excited modes in 
$T$, $S$ and $A_M$ have masses of the order or heavier than $M_{\rm GUT}$, 
so that they are decoupled in the low energy physics on the domain 
walls \cite{1stpaper}.

\section{Moduli stabilization}
\label{sec:moduli_stabilization}
So far, we have considered $V_1 = 0$ where we have degeneracy 
of symmetric and symmetry-breaking vacua.
It yields  undesirable massless scalar fields of $t_{3,2}$ ($s_{3,2}$) 
and light off-diagonal vector fields $b_\mu$ for $\Omega L \ll 1$. 
From now on we incorporate the moduli-stabilizating potential 
$V_1$ to make all the undesirable massless or light fields 
sufficiently massive. 
We first observe that the  moduli-stabilizing potential $V_1$ 
gives the following different values at the discrete degenerate vacua 
of the previous section
\begin{equation}
V_1(\left<5,0\right>) 
< V_1(\left<3,2\right>) < V_1(\left<4,1\right>). 
\label{eq:vacuum_ordering}
\end{equation}
Since they are the energy densities of the corresponding 
vacua, the energy densities of domain wall configurations in regions 
far away from domain walls can be approximated by these vacuum 
energy densities. 
Our boundary condition (\ref{eq:boundary_cond}) implies that 
the lowest energy vacuum $\left<5,0\right>$ should be chosen 
in the bulk. 
If there is a region of other vacuum between well-separated domain 
walls, the region should shrink because of higher vacuum energy density, 
resulting in a confining force between the split domain walls. 
This tendency persists till separated domain walls reach as close as 
the width of the walls. 
For small separation between domain walls, we find a residual repulsive 
force which decays exponentially at the wall width (quite common to 
interactions between solitons). 
Hence one can expect that there may be stable configuration with 
separated walls. 
Since the highest energy $\left<4,1\right>$ vacuum is energetically 
unfavorable, we exclude configurations containing the vacuum 
$\left<4,1\right>$ out of ten possible patterns of wall 
configurations. 
Then we are left with only three possibilities: 3-2 splitting, 
2-1-2 splitting or no splitting. 

\begin{figure}[t]
\begin{center}
\includegraphics[width=8.5cm]{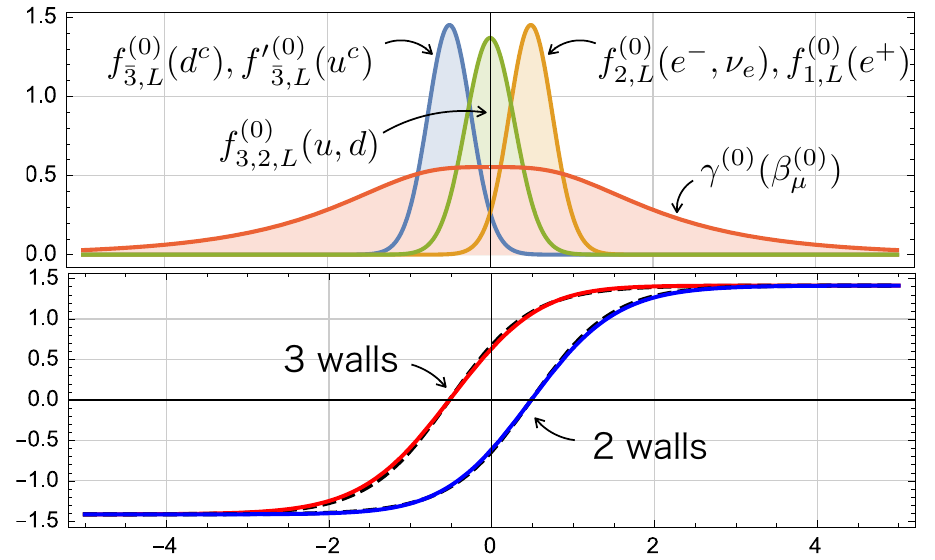}
\end{center}
\caption{(bottom) The solid lines show the numerical solution for the 
diagonal elements of $T$ for $v = \sqrt2$, 
$\lambda = \Omega = 1$, $\mu = 0.81$ and $\alpha = 1.1$. 
The broken curves stand for the analytic solution 
(\ref{eq:gensol3}) for $V_1=0$ with the same separation $L=1$. 
(top)
The mode functions of the fermions and massive gauge boson with $h=\tilde h=10$. 
The shared horizontal axis is $y$.}
\label{fig:wall_sol}
\end{figure}

\begin{figure}[t]
\begin{center}
\includegraphics[width=8.5cm]{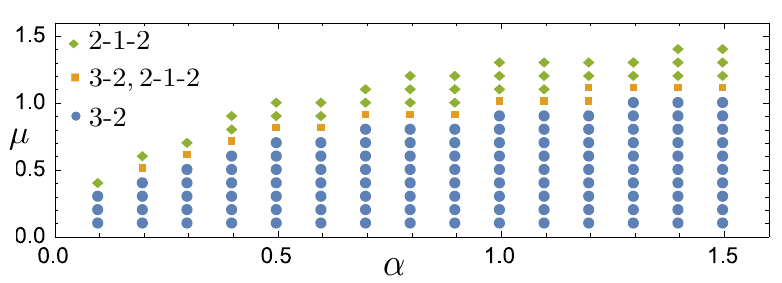}
\end{center}
\caption{
Minima in the $\alpha$-$\mu$ space. 
The 3-2 (2-1-2) splitting is the global minimum for the blue 
circle (green diamond). Both of them exist as 
global and local minima for the orange square ($v = \sqrt2$, 
$\lambda = \Omega = 1$).}
\label{fig:survey}
\end{figure}

To figure out which is the true ground state configuration, 
we survey the parameter space $\alpha$-$\mu$ by using the gradient flow 
equation \cite{Manton:2004tk}
\begin{equation}
\frac{\partial \phi(y,t')}{\partial t'}
=-\frac{\delta S_{\rm E}}{\delta \phi(y,t')}.
\end{equation}
Here, $\phi$ stands for the scalar fields and we introduce a 
fictitious flow-time $t'$ and let the field 
configuration relax  toward the local minimum 
of the Euclidean action $S_{\rm E}$ as $t'\to\infty$. 
To find global minima, we use initial configurations in 
(\ref{eq:gensol3}) and solve the gradient flow equation with 
randomly chosen $Y$ for each point in the $\alpha$-$\mu$ space 
repeatedly.
We confirm 
that there exists a large region in the $\alpha$-$\mu$ parameter 
space where the 3-2 splitting is the global minimum, see Fig.~\ref{fig:survey}. 
Our resut shows that the attractive force given by the higher 
vacuum energy of the $\left<3,2\right>$ vacuum is balanced against an 
exponentially decaying repulsive force near domain walls.  
Taking all the parameters with mass dimension in $V_{0}$ and $V_{1}$ 
to be the same order of magnitude, and identify them with $M_{\rm GUT}$, 
the 3-2 walls are stabilized at the separation 
\begin{equation}
L \sim M_{\rm GUT}^{-1}, 
\end{equation}as illustrated in 
Fig.~\ref{fig:wall_sol}. 
Thus, $V_1$ lifts the moduli in $t_{2,3}$ ($s_{2,3}$) and 
gives the mass 
\begin{equation}
\tilde \mu_0 \sim M_{\rm GUT}
\end{equation} 
to $b_\mu$.
At the same time, we find that all the undesirable 
moduli fields are stabilized to have masses of order 
$M_{\rm GUT}$.

\section{Quarks and Leptons}
\label{sec:quark_lepton}
The Yukawa terms for $\Psi_{\bar 5,10}$ and $T$ provides
massless chiral fermions around the domain walls. 
The fermion localization depends only on the solution $T$ 
which we can obtain reliably by numerically solving gradient flow 
even when $V_1 \neq 0$. 
As shown in Fig.~\ref{fig:wall_sol}, difference between the 
numerical solution (solid curves) for $V_1 \neq 0$ 
and the analytic solution (dashed curves) given in Eq.~(\ref{eq:gensol3}) 
is tiny. 
Thus, we can make use of (\ref{eq:gensol3}) with 
$Y = {\rm diag}({\cal Y}_3,{\cal Y}_3,{\cal Y}_3,{\cal Y}_2,{\cal Y}_2)$ 
in the following.
Let us decompose the fermions as 
$\bar5 \to (\bar 3,1)_{1/3} \oplus (1,2)_{-1/2}$  
and  $10 \to (\bar 3,1)_{-2/3} 
\oplus (3,2)_{1/6} \oplus (1,1)_1$ 
\be \label{eq:fdecom}
\Psi_{\bar 5} 
= \left(\begin{
matrix}
\Psi_{\bar 3} \\
\Psi_{2} 
\end{
matrix}
\right)\,,
\quad
\Psi_{10} = \left(\begin{
matrix}
\Psi_{\bar3}' & \Psi_{3,2} \\
-\Psi_{3,2}^t & \Psi_1
\end{
matrix}
\right)\,.
\ee

We expand fermions into modes such as 
\begin{equation}
\Psi_{\bar 3} = \sum_n \left[
f_{\bar 3,L}^{(n)} (y)\psi_{\bar 3,L}^{(n)}(x) 
+ f_{\bar 3,R}^{(n)} (y)\psi_{\bar 3,R}^{(n)}(x)
\right], 
\end{equation}
where effective fields $\psi_{L,R}(x)$ are chiral 
fermions in $3+1$ dimensions defined by 
\begin{equation}
\psi_L = P_-\psi, \quad 
\psi_R = P_+\psi, 
\end{equation}
with $P_\pm \equiv (1\pm \gamma^5)/2$. 
Furthermore, 
four-dimensional gamma matrices $\gamma^\mu$ are
defined from the five-dimensional gamma matrices $\Gamma^M$ as 
\begin{equation}
\Gamma^{\mu} = \gamma^\mu, \quad 
\Gamma^{5}= i\gamma^5, \quad 
\gamma^5 = i\gamma^0\gamma^1\gamma^2\gamma^3.
\end{equation}
The other components can be expanded similarly.
Plugging these into linearized field equations for 
$\Psi_{\bar 5, 10}$, we find equations for mode 
functions  
\begin{equation}
{\cal H}_{X} f_{X}^{(n)}(y) = m_{X}^{(n)}{}^2 f_{X}^{(n)}(y), 
\end{equation}
where the subscript $X$ distinguishes the components 
in Eq.~(\ref{eq:fdecom}).
The Hamiltonian for the left component ${\cal H}_{X,L}$ and the 
right component ${\cal H}_{X,R}$ are 
\begin{equation}
{\cal H}_{X,L} = Q_{X}^\dagger Q_{X}, \quad 
{\cal H}_{X,R} =Q_{X} Q_{X}^\dagger, 
\end{equation}
with 
\begin{equation}
Q_{X} = \p_y + h_X(y), \quad Q_{X}^\dagger = -\p_y + h_X(y).
\end{equation}

The $y$-dependent effective Yukawa couplings are approximated 
by $h_X = h \tau_3$ for $\Psi_{\bar 3}$, $\tilde h \tau_3$ for 
$\Psi_{\bar 3}'$, $h \tau_2$ for $\Psi_2$,  $\tilde h \tau_2$ for $\Psi_1$, 
and $\tilde h (\tau_3+\tau_2)/2$ for $\Psi_{3,2}$ using 
$\tau_3(y), \tau_2(y)$ defined in Eq.(\ref{eq:tau_function}).
The mode functions $f^{(0)}_X$ for zero modes are the 
kernels of $Q_X$ and $Q_X^\dagger$. 
When $h > 0$ and $\tilde h > 0$, the normalizable zero modes 
appear only in the left-handed components as 
\be
f_{\bar 3,L}^{(0)} =  N_{\bar 3}\left[\cosh \Omega 
(y - {\cal Y}_3)\right]^{-\frac{h v}{\Omega}}, 
\ee
with the normalization constant $N_{\bar 3}$. 
Similarly zero modes appear only for left-handed components 
$\psi_{\bar 3,L}'$, $\psi_{2,L}$ and $\psi_{1,L}$, whose mode 
functions are obtained by replacing $h$ with $\tilde h$, 
and ${\cal Y}_3$ with ${\cal Y}_2$. 
The mode functions 
for $\psi_{3,2,L}$ are given as 
\be
\!\!\!\!\!\!\!\!\!\!\!\!\!\!
f_{3,2,L}^{(0)} =  N_{3,2}
\left[\cosh \Omega (y - {\cal Y}_3)\cosh \Omega 
(y - {\cal Y}_2)\right]^{-\frac{\tilde h v}{2\Omega}}\,,
\ee
with the normalization constant $N_{3,2}$.

We identify one generation of quarks and leptons as usual 
\begin{equation}
\psi_{\bar3,L}^{(0)} \to d^c_\alpha, \quad 
\psi_{\bar 3,L}^{(0)}{}' \to u^c_\alpha, \quad 
\psi_{2,L}^{(0)} \to l_a = (e^-,\nu_e), 
\end{equation}
\begin{equation}
\psi_{1,L}^{(0)} \to e^+, \quad 
\psi_{3,2,L}^{(0)} \to q_{\alpha a} = (u_\alpha,\ d_\alpha). 
\end{equation}
We show typical mode functions in Fig.~\ref{fig:wall_sol},
where we observe 
that quarks and leptons are localized around the 
walls associated to the gauge fields with which they interact.

Maintaining charge universality has been difficult for 
localized gauge fields \cite{Rubakov2, Dubovsky:2001pe}. 
In our model, the charge universality 
holds exactly since the $3+1$ dimensional
SM gauge symmetry is fully maintained. 
In other words, 
the overlap integrals with the SM gauge fields 
and quarks (leptons) are independent of wall positions, since mode functions of massless gauge fields are  flat 
$v_\alpha^{(0)}/\sigma_\alpha = 1$ as in Eq.(\ref{eq:flat_profile}).

\section{Baryon number violating process}
\label{sec:baryon_number_viol}
The massive gauge boson $\beta_\mu^{(0)}$ with mass $\tilde 
\mu_0 \sim M_{\rm GUT}$ plays the same role as $X,Y$ gauge 
bosons in the usual $3+1$ dimensional GUT, where its exchange 
is suppressed by the large mass $M_{\rm GUT}$. 
When the symmetry breaking occurs by the wall splitting, 
however, their coupling to quarks and leptons can have 
significant additional suppression.
This suppression results from the small overlap of mode 
functions caused by two factors: the quark and lepton mode 
functions can be displaced from each other \cite{ArkaniHamed:1999dc}, 
and their mode functions are sharply peaked when their Yukawa 
couplings are large. 
The relevant triple couplings for the proton decay can be 
found in the kinetic terms of $\Psi_{\bar 5,10}$ as 
\be
{\cal L}_{\rm PD} &=& 
g H_{{\rm lq1}} \bar d^{\rm c}\gamma^\mu \beta^{(0)*}_\mu l
+ g H_{\rm dq}
\Tr_3\!\! \left[\bar U^{\rm c}\gamma^\mu \beta_\mu^{(0)} q^t \right]  \nonumber\\
&+&~ g H_{{\rm lq2}}
\Tr_2\!\!\left[\beta_\mu^{(0)t} \bar q \gamma^\mu E^+ \right] 
+ \text{h.c.}\,.
\ee
Here, we use the canonical normalization for the massive gauge 
bosons by redefining $\beta_\mu^{(0)} 
\to g \beta_\mu^{(0)}$, and identify fermion zero modes as
\begin{equation}
(E^+)_{ab} = \epsilon_{ab}e^+, \quad 
(U^{\rm c})_{ab} = \epsilon_{abc}u^{\rm c}_c.
\end{equation}
The triple couplings depend on the separation $L$ between the 
domain walls and are given by the
overlap integrals 
\begin{equation}
H_Z(L) = \int _{-\infty}^\infty dy\, r_Z, 
\end{equation} 
for the channel $Z={\rm lq1},\,{\rm lq2},\, {\rm dq}$ with 
the integrand given by
\begin{equation}
r_{{\rm lq1}} =
\frac{f_{\bar3,L}^{(0)}f_{2,L}^{(0)}\gamma^{(0)}}{\sigma_+},
\end{equation}
\begin{equation}
r_{{\rm lq2}} =
\frac{2 f_{1,L}^{(0)} f_{3,2,L}^{(0)} \gamma^{(0)}}{\sigma_+}, 
\end{equation}
\begin{equation}
r_{\rm dq} =
\frac{2f_{\bar3,L}'^{(0)} f_{3,2,L}^{(0)} \gamma^{(0)}}{\sigma_+}.
\end{equation}
\begin{figure}[t]
\begin{center}
\includegraphics[width=8.5cm]{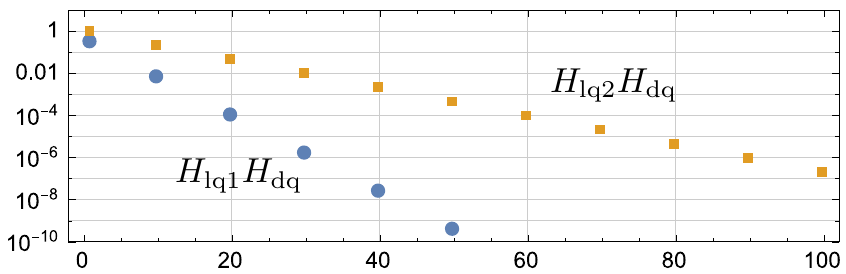}
\caption{Suppression factor $H_{\rm lq1}H_{\rm dq}$ (blue circle) and 
$H_{\rm lq2}H_{\rm dq}$ (orange square) as a function of $h = \tilde h$.
The parameters are $v = \sqrt2$ and $\Omega = \lambda = 1$ and we set $L=1$.}
\label{fig:proton_decay}
\end{center}
\end{figure}
The effective four-fermi coupling ${\cal G}_{ZZ'}$ for the process $Z \to Z'$ 
is given by 
\be
{\cal G}_{ZZ'} = \frac{H_Z H_{Z'} g^2}  {\tilde \mu_0^2} \,.
\ee
The suppression factors $H_{Z}$, and $H_{Z'}$ compared to the 
standard GUT is calculable in our model, since the mode function 
$\gamma^{(0)}(y)$, shown in Fig.~\ref{fig:wall_sol}, is reliably obtained 
unlike in the previous model \cite{Dvali:1996xe} 
based on an ambitious assumption of nonperturbative 
confinement in $4+1$ dimensions. 
On top of being a function of separation $L$ between walls, $H_{Z,Z'}$ 
also depends strongly on the magnitude of Yukawa couplings $h$ 
and $\tilde h$. 
The width of the mode function is inversely proportional to 
Yukawa couplings $h$ and $\tilde h$. 
Hence the large Yukwa coupling suppresses the overlap 
exponentially $H_Z \sim e^{-v h L}$. 
We plot 
$H_{\rm lq1}H_{\rm dq}$ and $H_{\rm lq2}H_{\rm dq}$ 
in Fig.~\ref{fig:proton_decay}, where we see that they are suppressed by more 
than $10^{-3}$ and $10^{-9}$, respectively, with the Yukawa 
coupling of order $50$. 
We note that there is no constraints from experimental 
data for these 
large Yukawa couplings between fermions and scalars 
forming the domain wall.

Finally let us consider the baryon number violating process 
mediated by the off-diagonal component $\tilde t$ of scalar field 
$T$ through the Yukawa couplings. 
Although it is difficult to solve the Schr\"odinger problem 
for $\tilde t$ \cite{1stpaper} analytically, it is still possible to get 
qualitative features. 
Since the zero modes of $\tilde t$ are absorbed by $b_\mu$, 
the lightest mode of $\tilde t$ should have mass of order 
$M_{\rm GUT}$. 
In addition, since the potential of the Schr\"odinger problem 
should make the mode functions localized around the walls, 
similarly to zero mode case, the overlaps of mode functions 
are suppressed similarly to the vector case. 
Hence we expect that $T$-mediated proton decay enjoys a 
suppression factor of the same order as that of the massive 
vector bosons.

\section{Conclusion}
\label{sec:conclusion}
We propose a $4+1$ dimensional model which unifies $SU(5)$ GUT 
and the brane world scenario. 
Our $3+1$ dimensional spacetime dynamically emerges with the 
symmetry breaking $SU(5)\to G_{\rm SM}$ together 
with one generation of the SM matter fields. 
We solve the gradient flow equation and confirm the 3-2 splitting configuration
is the global minimum in a large parameter region.
By applying the idea of the field-dependent gauge kinetic 
function \cite{Ohta,Us1,Us2} to our model, we solve the long-standing 
difficulties of the localization of massless gauge fields 
and charge universality.
All the undesirable moduli are stabilized. Furthermore, the proton 
decay can be exponentially suppressed.

Our model is an effective theory valid up to the GUT energy 
scale which is characterized by the inverse of the width of the 
domain wall. 
Above this GUT scale, we need to take account of effects 
of strong gauge dynamics, since our gauge field localization mechnism 
is based on a semi-classical representation of confinement in the bulk 
away from domain walls.
Below the GUT scale, the SM gauge couplings run following the usual 
renormalization group flow dictated by the low-energy effective 
Lagrangian in four dimensions. 
However, above the GUT scale, the effects of strong gauge dynamics 
should contribute to the running and change its behavior, in addition 
to the fact that the theory becomes five-dimensional. 
The quantum theory of our model above the GUT scale is an 
interesting subtle issue to study as a future work.

We have not yet included the SM Higgs field \cite{Maru:2001ch,Haba:2002if} and the second and 
higher generations, but our framework can easily incorporate 
the former similarly to Ref.~\cite{Volkas} and the latter with 
the mass hierarchy in the spirit of Ref.~\cite{ArkaniHamed:1999dc, Dvali2}.  
Furthermore, our model can be extended to other GUT gauge 
groups such as $SO(10)$.
Supersymmetry and/or warped spacetime with gravity can also be 
included without serious difficulties. 
Since our model has strong resemblance to D-branes in superstring 
theory, we hope that our field theoretical model can give some 
hints for simple construction\sout{s} of SM by D-branes.
\vspace{8mm}
\acknowledgements

F.\ B.\ was an international research fellow of the Japan Society 
for the Promotion of Science, and was supported by Grant-in-Aid 
for JSPS Fellows, Grant Number 26004750. 
This work is also supported in part 
by the Ministry of Education,
Culture, Sports, Science (MEXT)-Supported Program for the Strategic
Research Foundation at Private Universities ``Topological Science''
(Grant No.~S1511006), 
by the Japan Society for the 
Promotion of Science (JSPS) 
Grant-in-Aid for Scientific Research
(KAKENHI) Grant Numbers 25400280 (M.A.),  
26800119 and 16H03984 (M.\ E.),
25400241 (N.\ S.), by the Albert Einstein Centre for Gravitation and Astrophysics financed by the Czech Science Agency Grant No. 14-37086G (F.\ B.) and by the program of Czech Ministry of Education Youth and Sports INTEREXCELLENCE Grant number LTT17018 (F.\ B.).


\end{document}